\documentclass[aps,prx,twocolumn,groupedaddress,amsmath,amssymb,superscriptaddress]{revtex4}
\usepackage{graphicx}
\usepackage{subfigure}
\usepackage{hyperref}
\usepackage{epstopdf}
\usepackage{color}
\usepackage{multirow}
\usepackage{amsmath}  
\usepackage{amsfonts} 

\newcommand{\bq}{\begin{equation}}
\newcommand{\eq}{\end{equation}}
\newcommand{\bqa}{\begin{eqnarray}}
\newcommand{\eqa}{\end{eqnarray}}
\newcommand{\nn}{\nonumber \\}

\def\be     {\begin{equation}}
\def\ee     {\end{equation}}
\def\bea        {\begin{eqnarray}}
\def\eea        {\end{eqnarray}}
\def\bnn    {\begin{eqnarray*}}
\def\enn    {\end{eqnarray*}}

\begin{document}


\title{Entangled polymer complex as Higgs phenomena}

\author{Ki-Seok Kim}
\affiliation{Department of Physics, POSTECH, Pohang, Gyeongbuk 790-784, South Korea}
\author{Sandipan Dutta}
\affiliation{Asia-Pacific Center for Theoretical Physics, Pohang, Gyeongbuk 790-784, South Korea}
\author{YongSeok Jho}
\email{ysjho@apctp.org}
\affiliation{Asia-Pacific Center for Theoretical Physics, Pohang, Gyeongbuk 790-784, South Korea}
\affiliation{Department of Physics, POSTECH, Pohang, Gyeongbuk 790-784, South Korea}

\date{\today}

\begin{abstract}
We derive an effective Maxwell-London equation for entangled polymer complex under the topological constraint, borrowing the theoretical framework from the topological field theory.
We find that the transverse current flux of a test polymer chain, surrounded with the entangled chains, decays exponentially from its average position with finite penetration depth, which is analogous to the magnetic-field decay in a superconductor (SC). Like the mass acquirement of photons in SC is the origin of the magnetic-field decay, the polymer earns uncrossable intersections along the chain due to the preserved linking number, which restricts the deviation of the transverse polymer current in the normal direction.
%
%
Interestingly, this picture is well incorporated within the most successful phenomenological theory of the so called tube model, of which researchers have long pursued its microscopic origin. The correspondence of our equation of motion to the tube model claims that the confining tube potential is a consequence of the topological constraint (linking number). The tube radius is attributed to the decay length, and increasing the retracting force at intersections or increasing the number of intersections (linking number), the tube becomes narrow and tighter. It further shows that the probability of the tube leakage decays exponentially with the decay length of the tube radius.

\end{abstract}

\maketitle

\underline{\it{Introduction}:}
It is an intriguing nature of the physics that more than two completely different systems are described within the same mathematical framework. The Chern-Simons theory which was first introduced in string theory by Witten at 1980's~\cite{witten1989quantum}, turns out to be applicable to the cutting edge problems in condensed matter, such as fractional quantum Hall effect~\cite{zhang1992chern} and topological insulator and superconductor~\cite{kane2005z}. Since Edwards first introduced the topological constraint~\cite{edwards1967statistical}, \textit{i.e.} linking number, to the partition function of the polymer melts, researchers noticed the potential usefulness to apply the Chern-Simons theory to polymer problems~\cite{tanaka1982gauge,ferrari1998chern,otto1996entangled,ferrari2002statistical}. In those works, the knotted electron's movement is translated to the intersection (entanglement) of polymers. In spite of its mathematical exactness, these approaches were not very successful in polymer entanglement unlike its huge achievement in other fields of physics~\cite{everaers1999entanglement,mcleish2002tube}. Main reasons are that the model is still abstract, and the information of microstates is not feasible under current experimental techniques~\cite{everaers1999entanglement,mcleish2002tube}.

Surprisingly, long polymer melts, which is one of the most complicated system in polymer physics, is well described by a single body interpretation, \textit{i.e.} tube model~\cite{edwards1967statistical2,doi1986theory,de1971reptation}. In addition to many rheological experimental supports~\cite{fetters1993rheological,watanabe2002dielectric,wang2013new,mcleish2002tube}, recent experiments and simulations verified the validity of the model finding the linear confinement potential around semi-flexible polymers~\cite{robertson2007direct,wang2010confining,glaser2010tube,zhou2006direct}. The strength of this intuitive model lies in its universality spanning over the intermediate and long time scale regardless of the detailed chemistry or the structure of the polymer, as long as the polymer is long enough so that we can treat it as a coarse grained random chain with the stepsize of random work much exceeding the local length scale of the polymer. Since this phenomenological model earned a great success, researchers have sought its microscopic origin~\cite{kuzuu1980rheology,zhou2006direct,everaers2004rheology}. Recent theoretical achievements in this effort have focused on the Langevin approach of statistical mechanics~\cite{sussman2011microscopic,sussman2012microscopic}.
%
%

These two approaches, Chern-Simons theory and the tube model, look completely different. However, we believe that the phenomenological theory should be understood from the microscopic theory. The goal of the current work is to pursue this connection.

As Edwards claimed in his pioneering work, polymer conformation can be translated as a classical particle motion~\cite{edwards1965statistical,doi1986theory}. In this analogy, time corresponds to the displacement along the chain, and the trajectory of the particle motion is associated with the contour of the polymer. Within this argument, the diffusive motion of a classical particle can be interpreted as a Gaussian distribution of a random polymer chain. To include the entanglement effect, he plugged the topological constraint into the partition function~\cite{edwards1967statistical}. This entangled polymer system has a similar mathematical structure to the trajectory of a classical particle under a magnetic field following the Biot-Savart law, which allows us to analyze the polymer entanglement in favor of the magnetic-field induction under the constraint of integer linking number. This leads us to construct BF theory~\cite{cho2011topological} that avoids self-linking contrary to Chern-Simons theory.

We agree that it is extremely difficult to solve such complicate equations obtained from Chern-Simons theory exactly and provide the correspondence of parameters between the microscopic and phenomenological theories~\cite{everaers1999entanglement,mcleish2002tube}. In this respect we adopt a new strategy to make a link between the phenomenological tube model and the microscopic BF theory. Inspired by previous works~\cite{tanaka1982gauge,ferrari1998chern,otto1996entangled,ferrari2002statistical}, we derive an effective Maxwell-London equation for the dynamics of entangled polymer melts, starting from the topological BF theory. We interpret the characteristic features of the tube model with this Maxwell-London equation, which leads us to identify the dynamics of entangled polymers with the Higgs phenomena.

\begin{table*}[!htb]
  \begin{tabular}{|c||c|c|}
    \hline
     \textbf{System} & \textbf{Superconductor (SC)} & \textbf{Entangled polymer} \\ \hline \hline
     \multirow{2}{*}{Medium} & SC vacuum & Entangled polymer complex \\ \cline{2-3}
     & SC boundary & Local averaged position of a chain \\ \hline
     \multirow{3}{*}{Degrees of freedom}& Current & Tangential flux of a chain \\ \cline{2-3}
     & External magnetic field & Test chain \\ \cline{2-3}
     & Magnetic field induction & Flux of entangled chains \\ \hline
     Topology constraint & Flux quantization & Preserving linking number \\ \hline
     \multirow{3}{*}{Physics}& Photon mass & Retracting force at intersections \\ \cline{2-3}
     & Magnetic field penetration & Diffusion of tangential flux \\ \cline{2-3}
     & Decay length & Tube radius \\ \hline
  \end{tabular}
  \caption{An analogy between polymer entanglement (tube model) and superconductor}
  \label{table:table1}
\end{table*}

\begin{figure}[ht]
    \includegraphics[width=8cm]{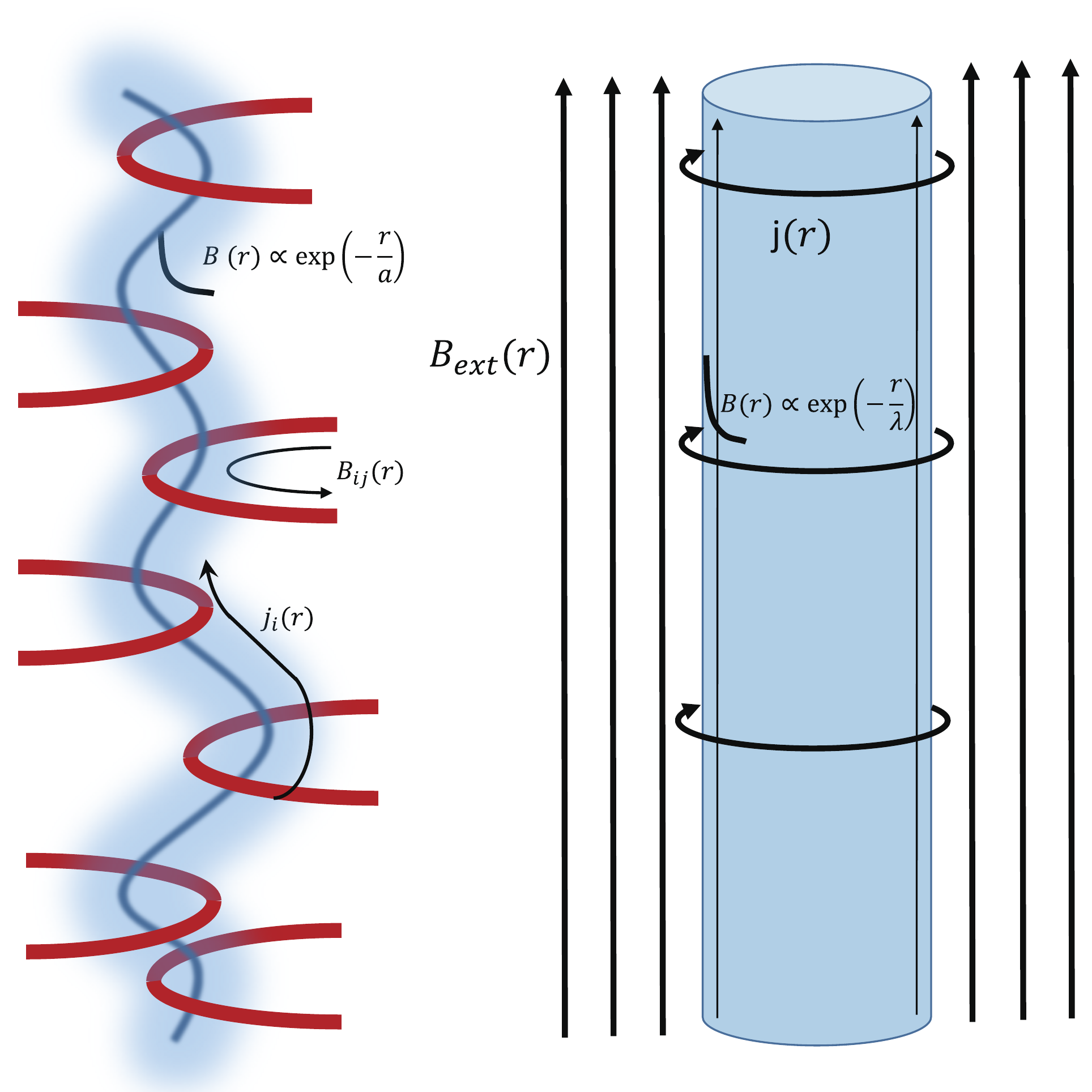}
    \caption{Schematic illustration of topological constraint of polymer melts. We assume that the number of entanglement is preserved during time scale we are interested in.  }
    \label{fig:fig1}
\end{figure}

\underline{{Topological field theory}:}
The polymer density and current for $N$ chains of length $L$ are given by
\begin{align}
 \rho(\mathbf{r}) & = \sum_{i = 1}^{N} \int_0^L ds\delta(\mathbf{x}_i(s)-\mathbf{r}) \\
 \mathbf{j}(\mathbf{r}) & = \sum_{i = 1}^{N} \int_0^L ds\dot{\mathbf{x}}(s)\delta(\mathbf{x}_i(s)-\mathbf{r}) .
 \label{eq1}
\end{align}
Introducing an artificial electromagnetic vector potential $\mathbf{A}_I(\mathbf{r})$ given by other polymers except for the I$^{th}$
\begin{align}
 \nabla\times\mathbf{A}_I(\mathbf{r}) = \sum_{j\neq I}\mathbf{j}_j(\mathbf{r}) ,
 \label{eq2}
\end{align}
where the most general solution to the Eq.~\eqref{eq2} is
\begin{align}
 \mathbf{A}_I(\mathbf{r}) & = \frac{1}{4\pi}\int \sum_{j\neq I}\mathbf{j}_j(\mathbf{r}^{\prime})\times\frac{\mathbf{r}-\mathbf{r}^{\prime}}
 {\vert\mathbf{r}-\mathbf{r}^{\prime}\vert^3}d\mathbf{r}^{\prime} ,
 \label{eq3}
\end{align}
it is convenient to express the winding number of the I$^{th}$ polymer with others as follows
\begin{equation}
 N_I = \int d\mathbf{r}\sum_I\mathbf{A}_I(\mathbf{r}) \cdot \mathbf{j}_I(\mathbf{r}) .
 \label{winding}
\end{equation}
If we consider only two polymer chains for simplicity, it is straightforward to see how Eq.~\eqref{winding} shows the winding number
\begin{align}
 N_{2} & = \int d\mathbf{r}\mathbf{A}_2(\mathbf{r}) \cdot \mathbf{j}_2(\mathbf{r}) \nonumber\\
 & = \int \frac{1}{4\pi}\int \mathbf{j}_2(\mathbf{r}^{\prime})\times\frac{\mathbf{r}-\mathbf{r}^{\prime}}{\vert\mathbf{r}-\mathbf{r}^{\prime}\vert^3}
 d\mathbf{r}^{\prime} \cdot \mathbf{j}_1(\mathbf{r})d\mathbf{r} \nonumber\\
 & = \frac{1}{4\pi}\oint\oint d\mathbf{r}^{\prime}(s)\times d\mathbf{r}^{\prime\prime}(s^{\prime})\cdot
 \frac{\mathbf{r}^{\prime}(s)-\mathbf{r}^{\prime\prime}(s^{\prime})}{\vert\mathbf{r}^{\prime}(s)-\mathbf{r}^{\prime\prime}(s^{\prime})\vert^3} .
\end{align}

Introducing this topological constraint into the partition function for the I$^{th}$ polymer, we obtain
\begin{widetext}
\bqa \mathcal{Z}_{I} &=& \sum_{N_{I} = 0}^{\infty} e^{- \mu_{I} N_{I}} \sum_{N_{Ij} = 0}^{\infty} \delta\Bigl(\sum_{j \not= I}^{N} N_{Ij} - N_{I}\Bigr) \frac{\Pi_{j \not= I}^{N} N_{Ij} !}{N_{I}!} \left\langle \delta\left(\nabla\times\mathbf{A}_I(\mathbf{r}) - \sum_{j\neq I}\mathbf{j}_j(\mathbf{r}) \right) \delta\biggl(N_I - \int d\mathbf{r}\sum_I\mathbf{A}_I(\mathbf{r}) \cdot \mathbf{j}_I(\mathbf{r})\biggr) \right\rangle \nn &=& \sum_{N_{I} = 0}^{\infty} e^{- \mu_{I} N_{I}} \sum_{N_{Ij} = 0}^{\infty} \delta\Bigl(\sum_{j \not= I}^{N} N_{Ij} - N_{I}\Bigr) \frac{\Pi_{j \not= I}^{N} N_{Ij} !}{N_{I}!} \int_{-\infty}^{\infty} d g_{I} e^{i g_{I} N_{I}} \int \Pi_{i = 1}^{N} D \mathbf{X}_{i}(s) D \mathbf{A}_{I}(\mathbf{r}) D \mathbf{C}_{I}(\mathbf{r}) \nn && \exp\biggl\{ - \Bigl( \int_{0}^{L} d s \sum_{i = 1}^{N} \frac{\mathcal{K}}{2} \mathbf{\dot{X}}_{i}^{2}(s) - i g_{I} \int d^{d} \mathbf{r} \mathbf{j}_{I}(\mathbf{r}) \cdot \mathbf{A}_{I}(\mathbf{r}) - i \int d^{d} \mathbf{r} \mathbf{C}_{I}(\mathbf{r}) \cdot \sum_{j \not= I}^{N} \mathbf{j}_{j}(\mathbf{r}) + i \int d^{d} \mathbf{r} \mathbf{C}_{I}(\mathbf{r}) \cdot [ \mathbf{\nabla} \times \mathbf{A}_{I}(\mathbf{r}) ] \Bigr) \biggr\} , \nn \eqa
\end{widetext}
where $g_{I}$ is a Lagrange multiplier field to impose the constraint $\delta\biggl(N_I - \int d\mathbf{r}\sum_I\mathbf{A}_I(\mathbf{r}) \cdot \mathbf{j}_I(\mathbf{r})\biggr)$ and $\mathbf{C}_{I}(\mathbf{r})$ is an auxiliary gauge field to impose the constraint $\delta\left(\nabla\times\mathbf{A}_I(\mathbf{r}) - \sum_{j\neq I}\mathbf{j}_j(\mathbf{r}) \right)$. $N_{Ij}$ is an integer to represent a Gaussian linking number between the I$^{th}$ polymer and $j_{th}$ one ($j \not= I$). $N_{I} = \sum_{j \not= I}^{N} N_{Ij}$ is an integer to express a total Gaussian linking number between the I$^{th}$ polymer and all others ($j \not= I$). $\mu_{I}$ is an energy cost per a linking event. Although Gaussian-fluctuation polymers have been assumed, the formal expression can be generalized to take their interactions.

It is straightforward to write down the canonical partition function for $N$ polymers, given by
\begin{widetext}
\bqa \mathcal{Z}_{N} &=& \sum_{N_{G} = 0}^{\infty} e^{- \mu_{G} N_{G}} \sum_{N_{I} = 0}^{\infty} \delta\Bigl(N_{G} - \sum_{I = 1}^{N} N_{I} \Bigr) \frac{\Pi_{I = 1}^{N} N_{I}!}{N_{G}!}
\sum_{N_{Ij} = 0}^{\infty} \delta\Bigl(N_{I} - \sum_{j \not= I}^{N} N_{Ij} \Bigr) \frac{\Pi_{j \not= I}^{N} N_{Ij} !}{N_{I}!} \int_{-\infty}^{\infty} \Pi_{I = 1}^{N} d g_{I} e^{i \sum_{I = 1}^{N} g_{I} N_{I}} \nn && \int \Pi_{I = 1}^{N} D \mathbf{X}_{I}(s) D \mathbf{A}_{I}(\mathbf{r}) D \mathbf{C}_{I}(\mathbf{r}) \exp\biggl\{ - \sum_{I = 1}^{N} \Bigl( \int_{0}^{L} d s \frac{\mathcal{K}}{2} \mathbf{\dot{X}}_{I}^{2}(s) - \frac{i}{2} \int d^{d} \mathbf{r} g_{I} \mathbf{j}_{I}(\mathbf{r}) \cdot \mathbf{A}_{I}(\mathbf{r}) - i \int d^{d} \mathbf{r} \mathbf{C}_{I}(\mathbf{r}) \cdot \sum_{j \not= I}^{N} \mathbf{j}_{j}(\mathbf{r}) \nn && + i \int d^{d} \mathbf{r} \mathbf{C}_{I}(\mathbf{r}) \cdot [ \mathbf{\nabla} \times \mathbf{A}_{I}(\mathbf{r}) ] \Bigr) \biggr\} , \eqa
\end{widetext}
where $N_{G} = \sum_{I = 1}^{N} N_{I}$ is the total Gaussian linking number and $\mu_{G}$ is an energy cost per a linking event. The partition function may be more manageable in a form of the grand canonical partition function, given by \bqa && Z_{GC} = \sum_{N = 0}^{\infty} \frac{e^{- \mu N}}{N!} Z_{N} , \eqa where $\mu$ is a chemical potential for a polymer.

An essential feature in this formulation lies in the so called BF term, given by \bqa && \mathcal{S}_{BF} = i \int d^{d} \mathbf{r} \sum_{I = 1}^{N} \mathbf{C}_{I}(\mathbf{r}) \cdot [ \mathbf{\nabla} \times \mathbf{A}_{I}(\mathbf{r}) ] , \eqa which describes mutual entanglement of polymers. In addition to the BF term, the dynamics of the fluctuating polymers will give rise to the Maxwell dynamics for the emergent gauge fields of $\mathbf{A}_{I}(\mathbf{r})$ and $\mathbf{C}_{I}(\mathbf{r})$. This may be regarded as the effect of renormalization for the dynamics of photons from the dynamics of electrons in the theory of quantum electrodynamics~\cite{peskin1995}. As a result, we obtain the so called BF-Maxwell theory as an effective description for the dynamics of entangled polymers, given by
\begin{widetext}
\bqa && \mathcal{S}_{eff} = \int d^{d} \mathbf{r} \sum_{I = 1}^{N} \Bigl( \frac{1}{2e_{c}^{2}} [\mathbf{E}_{\mathbf{C}_{I}}^{2}(\mathbf{r}) - \mathbf{B}_{\mathbf{C}_{I}}^{2}(\mathbf{r})] + \frac{1}{2e_{a}^{2}} [\mathbf{E}_{\mathbf{A}_{I}}^{2}(\mathbf{r}) - \mathbf{B}_{\mathbf{A}_{I}}^{2}(\mathbf{r})] + i \mathbf{C}_{I}(\mathbf{r}) \cdot [ \mathbf{\nabla} \times \mathbf{A}_{I}(\mathbf{r}) ] \Bigr) , \eqa
\end{widetext}
where both effective coupling constants of $e_{c}^{2}$ and $e_{a}^{2}$ are proportional to $g_{I}^{2}$ since these contributions originate from the fluctuating polymers. Considering \bqa \partial_{t} & \longrightarrow & \int_{0}^{L} d s \frac{d}{d s} = \int_{0}^{L} d s \sum_{i = 1}^{N} \frac{d \mathbf{X}_{i}(s)}{d s} \cdot \frac{\partial}{\partial \mathbf{X}_{i}(s)} \nn & = & \int_{0}^{L} d s \sum_{i = 1}^{N} \int d^{d} \mathbf{r} \delta^{(3)}(\mathbf{r} - \mathbf{X}_{i}(s)) \mathbf{\dot{X}}_{i}(s) \cdot \mathbf{\nabla}_{\mathbf{r}} , \nonumber \eqa we obtain \bqa && \partial_{t} \longrightarrow \sum_{i = 1}^{N} \mathbf{j}_{i}(\mathbf{r}) \cdot \mathbf{\nabla}_{\mathbf{r}} . \nonumber \eqa Then, effective electric and magnetic fields are given by \bqa && \mathbf{E}_{\mathbf{C}_{I}}(\mathbf{r}) = - \partial_{t} \mathbf{C}_{I}(\mathbf{r}) \equiv - \sum_{i = 1}^{N} \mathbf{j}_{i}(\mathbf{r}) \cdot \mathbf{\nabla} \mathbf{C}_{I}(\mathbf{r}) , \nn && \mathbf{B}_{\mathbf{C}_{I}}(\mathbf{r}) = \mathbf{\nabla} \times \mathbf{C}_{I}(\mathbf{r}) \eqa for $\mathbf{C}_{I}(\mathbf{r})$ and \bqa && \mathbf{E}_{\mathbf{A}_{I}}(\mathbf{r}) \equiv - \partial_{t} \mathbf{A}_{I}(\mathbf{r}) = - \sum_{i = 1}^{N} \mathbf{j}_{i}(\mathbf{r}) \cdot \mathbf{\nabla} \mathbf{A}_{I}(\mathbf{r})  , \nn && \mathbf{B}_{\mathbf{A}_{I}}(\mathbf{r}) = \mathbf{\nabla} \times \mathbf{A}_{I}(\mathbf{r}), \eqa for $\mathbf{A}_{I}(\mathbf{r})$.
%
%

\underline{{Maxwell-London equation}:}
Applying the least action principle to the BF-Maxwell action, it is straightforward to derive the following equation of motion
\begin{widetext}
\bqa && - \frac{1}{e_{c}^{2}} \sum_{i = 1}^{N} \sum_{i' = 1}^{N} [\mathbf{j}_{i}(\mathbf{r}) \cdot \mathbf{\nabla}] [\mathbf{j}_{i'}(\mathbf{r}) \cdot \mathbf{\nabla}] \mathbf{C}_{I}(\mathbf{r}) + \frac{1}{e_{c}^{2}} \mathbf{\nabla}^{2} \mathbf{C}_{I}(\mathbf{r}) - \frac{1}{e_{c}^{2}} \mathbf{\nabla} \mathbf{\nabla} \cdot \mathbf{C}_{I}(\mathbf{r}) = - i \mathbf{\nabla} \times \mathbf{A}_{I}(\mathbf{r}) , \nn && - \frac{1}{e_{a}^{2}} \sum_{i = 1}^{N} \sum_{i' = 1}^{N} [\mathbf{j}_{i}(\mathbf{r}) \cdot \mathbf{\nabla}] [\mathbf{j}_{i'}(\mathbf{r}) \cdot \mathbf{\nabla}] \mathbf{A}_{I}(\mathbf{r}) + \frac{1}{e_{a}^{2}} \mathbf{\nabla}^{2} \mathbf{A}_{I}(\mathbf{r}) - \frac{1}{e_{a}^{2}} \mathbf{\nabla} \mathbf{\nabla} \cdot \mathbf{A}_{I}(\mathbf{r}) = - i \mathbf{\nabla} \times \mathbf{C}_{I}(\mathbf{r}) . \eqa
\end{widetext}
If we focus on the linear regime for gauge dynamics, we are allowed to neglect the electric-field term. The reason is that the polymer current can be expressed in terms of gauge potentials, referred to as constituent equations and giving rise to higher-order dynamics of gauge fields. This may be regarded to be nothing but the Ohm's law in metals although it is not clear how to construct precise constituent equations in the present situation, where the dynamic information of the entangled polymers should be incorporated.

Taking the Coulomb gauge of $\mathbf{\nabla} \cdot \mathbf{C}_{I}(\mathbf{r}) = 0$ and $\mathbf{\nabla} \cdot \mathbf{A}_{I}(\mathbf{r}) = 0$, we find the Maxwell-London equation
\bqa && \frac{1}{e_{c}^{2}} \mathbf{\nabla}^{2} \mathbf{B}_{\mathbf{C}_{I}} (\mathbf{r}) = e_{a}^{2} \mathbf{B}_{\mathbf{C}_{I}} (\mathbf{r}) , \nn && \frac{1}{e_{a}^{2}} \mathbf{\nabla}^{2} \mathbf{B}_{\mathbf{A}_{I}} (\mathbf{r}) = e_{c}^{2} \mathbf{B}_{\mathbf{A}_{I}} (\mathbf{r}) . \eqa Recall \bqa && \mathbf{B}_{\mathbf{A}_{I}} (\mathbf{r}) = \sum_{j \not= I}^{N} \mathbf{j}_{j}(\mathbf{r}) , ~~~~~ \mathbf{B}_{\mathbf{C}_{I}} (\mathbf{r}) = \frac{g_{I}}{2} \mathbf{j}_{I}(\mathbf{r}) . \eqa We write down the Maxwell-London equation in a more suggestive fashion \bqa \label{eq:london} && \left(D \mathbf{\nabla}^{2} - M^{2}(\mathbf{r}) \right) \mathbf{B} (\mathbf{r}) = 0 , \eqa where the diffusion coefficient of $D$ is introduced for physical interpretation. It is interesting to notice that $M(\mathbf{r})$ can be identified with an effective mass of photons inside SC that corresponds to an inverse of the penetration depth of magnetic fields. In the setup of the BF theory it is proportional to $g_{I}^{2}$, where $g_{I}$ is the coupling function to enforce the constraint of the Gaussian linking number. Within our semiclassical framework, it can be determined by the saddle-point approximation for the resulting free energy. This procedure may be one of the most complex parts in our analysis. In the present study we do not perform this saddle point analysis and take the mass parameter as a function of the distance away from the average position of a test polymer. In the language of polymers, the penetration depth is read as a decay length of the transverse distribution of the test polymer current from its local averaged position. The mass may be understood as a retracting force at the intersections along the contour. In SC, the dissipationless current is induced to cancel the magnetic field inside as seen in Fig.~\ref{fig:fig1}, leaving the shallow penetration depth at the surface. In polymer melts, entangled chains (with a test chain) restrict large deviations of the test chain segment in the normal direction. Thus, we conclude that the decay of the transverse polymer current in the normal direction is an essential character resulting from the topological constraint. This picture is well incorporated in the tube model. In comparison with the tube model, the backbone of the tube is a local average position, and the radius of the tube corresponds to the decay length (the penetration depth in SC). In this argument, the exponential decay of the chain leakage out of the tube is well understood.
\begin{figure}[t]
    \includegraphics[width=9cm]{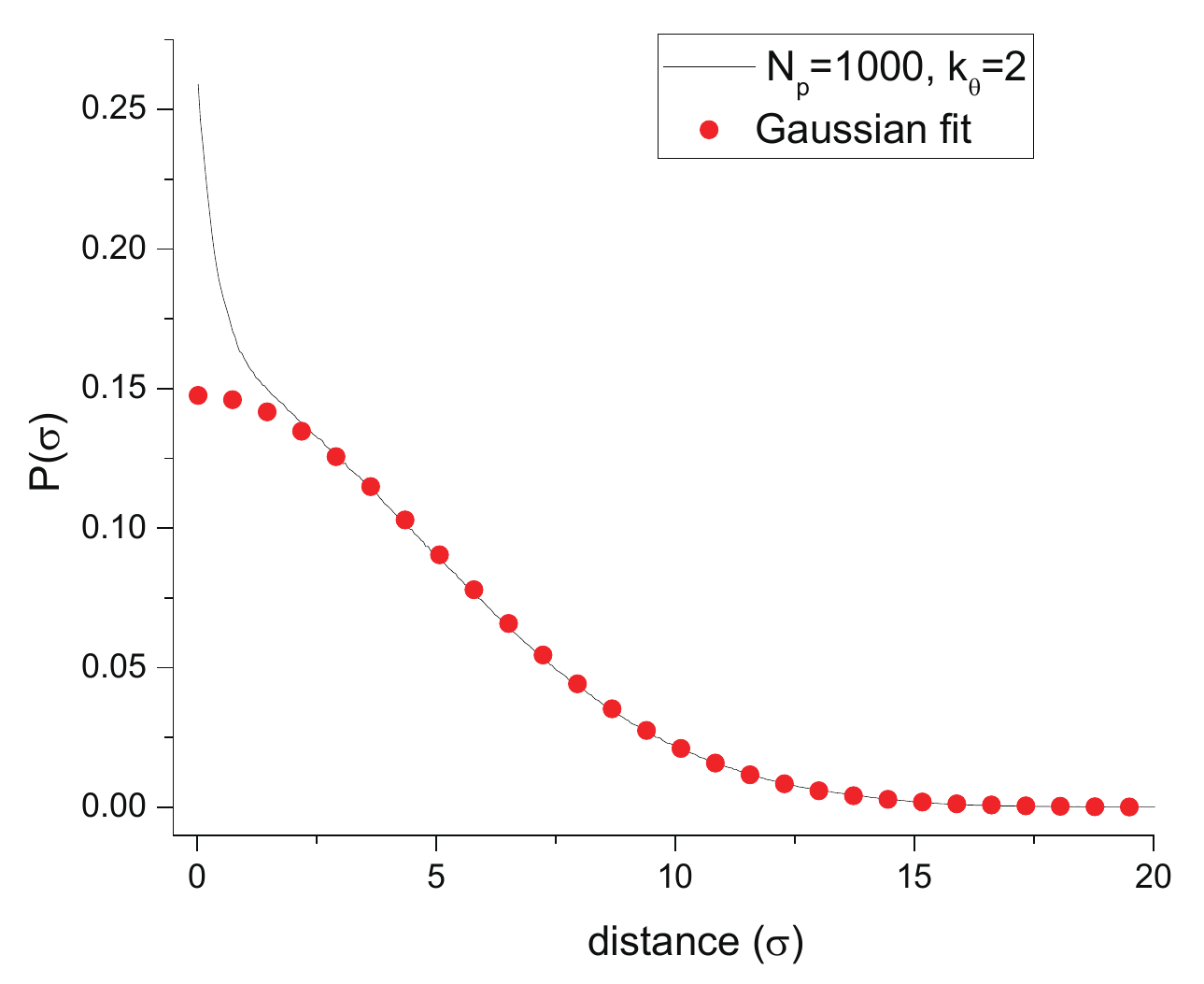}
    \caption{The transverse distribution of the chain is plotted at very low polymer density of $0.01$. The distribution is fitted with Gaussian perfectly except the center of local average position where the attraction of Lennard-Jones interaction affects.}
    \label{fig:fig2}
\end{figure}

\begin{figure}[t]
    \includegraphics[width=9cm]{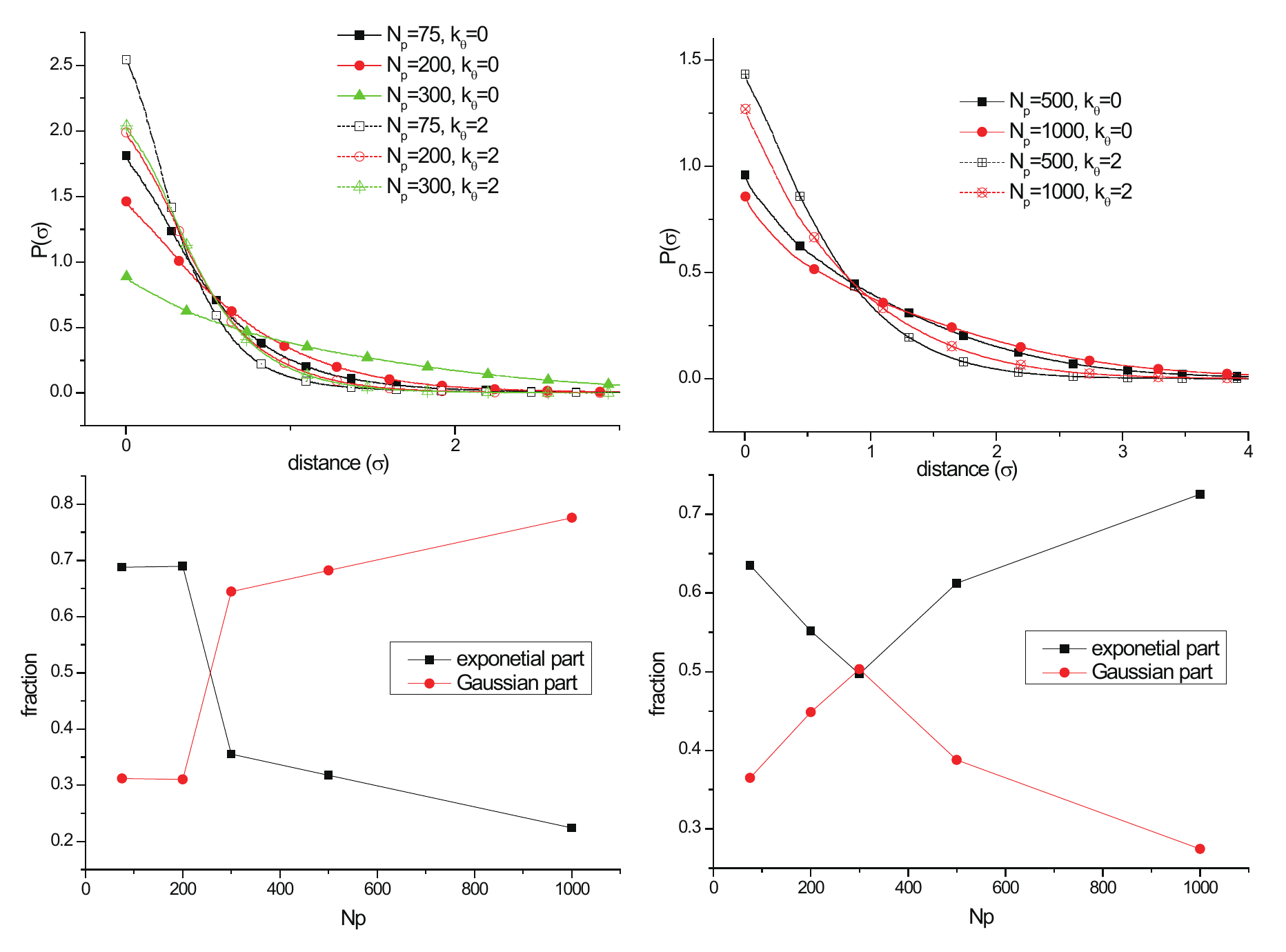}
    \caption{In (a), (b) the transverse distribution of the chain from the primitive path at different $N_p$, and $k_\theta$. In (c), (d) the areal fractions of Gaussian distribtuion and the exponential decaying function are plotted as function of $N_p$ when the transverse distributions are fitted with linear combination of Gaussian distribution and exponential decaying function.}
    \label{fig:fig3}
\end{figure}

\underline{Comparison with numerical results}

\underline{Method:}
The idea, that the transverse chain flux distribution is a linear combination of exponential decaying function and the Gaussian distribution, is tested by computer simulations. The linear potential to the normal direction of local average configuration will cause the exponenital decay of transverse current flux.
A computational time of the regular MD or MC simulations for entangled long polymer melts takes very long. Considering the relaxation time, which scales as about $N_p^4$ ( $N_p$ is polymerization degree.), we only are able to simulate short chain of polymer ($N_p \le 1000$).

Under this restriction, we performed molecular dynamics simulations based on the coarse-graining entanglement model developed by Kremer and Grest~\cite{Everaers06022004}.
Temperature of the system is kept to be $297 K$ using the Langevin equation.

Three interactions are considered to reflect the monomer size, chain connectivity, and the bending rigidity of the polymer.
The interaction between monomers is incorporated in a Lennard-Jones (LJ) potential,
\bqa U_{LJ} = \begin{cases} 4\epsilon \left\{ \left( \frac{\sigma_0}{r}\right)^{12} -\left( \frac{\sigma_0}{r}\right)^{6} + \frac{1}{4}\right\} & r<r_c \\
0& r\ge r_c \end{cases} \eqa
where, $r_c = 2^{\frac{1}{6}}\sigma_0$. $\sigma_0$ is a length unit in LJ model representing a diameter of monomer.

The chain is sustained by a FENE potential,
\bqa U_{FENE} = \begin{cases} -0.5 k R_0^2 \ln{\left( 1-(r/R_0)^2\right)}& r<R_0,
\\ \infty & r\ge R_0 \end{cases}\eqa
where $R_0 = 1.5 \sigma_0$. $\epsilon$ is an energy unit in LJ potential.
To adjust the stiffness of the chain, we add a bending potential,
\bqa U_{bend} = k_\theta \left(1-\cos{\theta} \right) \eqa
where, $k_\theta$ is the bending stiffness, and the angle $\theta$ is defined as $\cos{\theta_i} = \frac{\left(\vec{\mathbf{r}}_i - \vec{\mathbf{r}}_{i-1}\right)\cdot \left(\vec{\mathbf{r}}_{i+1} - \vec{\mathbf{r}}_{i}\right)}{\left|\left(\vec{\mathbf{r}}_i - \vec{\mathbf{r}}_{i-1}\right) \right|\left|\left(\vec{\mathbf{r}}_{i+1} - \vec{\mathbf{r}}_{i}\right) \right|}$.
When $k_\theta=0$, the chain is flexible. Chain becomes stiff increasing the bending stiffness, for example at $k_\theta=2$, the characerisitic ratio $C_\infty$ is about $3.4$ considered as semiflexible polymer. According to the reference~\cite{Everaers06022004}
$N_e^P$ is $65$ for $k_\theta = 0$, and $23$ for  $k_\theta = 2$.
As Kremer and Grest pointed out at $\epsilon = k_BT$, and $k = 30\frac{\epsilon}{\sigma_0^2}$, there is a huge energy barrier ($\simeq 70 k_BT$) for two chains to cross. This ensures the uncrossbility of the chain.

The time step is chosen as $\tau_0 = 0.01$. The friction coefficient for Gaussian noise is set $0.5 m/\tau_0$. To properly equilibrate the system, we adjust the simulation time scale to exceed the longest relaxation time (the disengagement time, $\tau_d = 4.5(N^3/N_e)\tau_0$). $10^8$ time steps are used for equilibration, and another $10^8$ time steps are used for the data production.

The transverse distribution of the chain from its local averaged position is measured within the constraint release time. For this purpose, we apply the direct tube sampling (DTS)~\cite{zhou2006direct}. At every $5 \times 10^6$ steps (which exceeds Rouse time, $\tau_R \simeq 1.5 N^2 \tau_0$) during production stage, we write out the conformation of polymer.
In DTS, we find the primitive paths of each conformation are saved during production stage. In order to keep the topological constraint preventing the reptation motion, disengagement or constraint release, we fix the ends of the polymers. Then, by cooling the temperature of the system to zero, we find out the energy minimized conformation of the polymer which is called primitive path. Then, we return back to the normal temperature and measure the deviation of the chain from the primitive path.

\underline{Results:} Linking number is preserved only for an intermediate time scale $\tau_d \gg t \gg \tau_{R}$ when the topological constraint affects the chain dynamics but the reptational diffusion, disengagement or constraint release are still suppressed. Incorporating the dynamical aspects of the entangled polymers into the static Maxwell-London equation for this regime, we propose an extended Maxwell-London equation
\bqa \label{eq:eqom} && \partial_{t} \mathbf{B} (\mathbf{r},t) = D \mathbf{\nabla}^{2} \mathbf{B} (\mathbf{r},t) - M^2(\mathbf{r}) \mathbf{B} (\mathbf{r},t) , \eqa
where
\bqa && \mathbf{B} (\mathbf{r},t) \propto \mathbf{j} (\mathbf{r},t) = \int_{0}^{L} d s \frac{d}{d s} \mathbf{X}(s,t) \delta^{(3)}(\mathbf{X}(s,t) - \mathbf{r}) . \nn \eqa
Within this time scale, the distribution of the intersection is not fully uniform over the chain, but localized. For the part of the chain far from the intersection, the diffusive motion would be dominant, \textit{i.e.} the polymer current of $\mathbf{J_I}$ will follow the Gaussian distribution, described by $\partial_{t} \mathbf{B} (\mathbf{r},t) = D \mathbf{\nabla}^{2} \mathbf{B} (\mathbf{r},t)$ especially, near $|\mathbf{r}| \approx 0$. In Fig.\ref{fig:fig1}, this idea is verified from the numerical simulation. At very low polymer density ($\rho=0.01$), the transverse distribution fits perfectly with a Gaussian function except at the vicinity of the local average position where the attraction of Lennard-Jones localizes the monomer distribution.
The diffusion time scale (or relaxation time at quenched intersections) is proportional to the chain length per the entanglement $\left(r_D \simeq \sqrt{N_e} \sigma_0\right)$ which is nothing but width of the Gaussian distribution. On the other hand, near the intersections the fluctuations of the test polymer will be suppresed. As a result, we expect a static solution in which the mass term is dominant, described by the previous Maxwell-London equation. The decay length $\lambda_d$ is about $\frac{\sqrt{D}}{M(\mathbf{r})}$. In a nutshell, the tube diameter can be justified as a dominant decaying length scale between $r_D$ and $\lambda_d$.

In our derivation, the mass term is proportional to the energy of linking events, and its spatial frequency (linking number), \textit{i.e.} the mass term is dominant when the interaction energy at intersection is strong and linking number is large.
One simple way to achieve both conditions is turning the bending energy up. In Fig.~\ref{fig:fig3}-(a),(b) we plot the transverse distribution of the chain as a normal distance $\sigma$ from its local averaged configuration varying $k_\theta$ (at $\rho = 0.85 \sigma_0^{-3}$). As we noted, increasing $k_\theta$, a chain transits from a flexible chain to a semi-flexible chain. In accordance, the linking number and interaction energy at the intersection increases. Indeed the transverse distribution of current decays faster for $k_{\theta}=2$ regardless $N_p$ (Fig.~\ref{fig:fig3}-(a),(b)).

For each distribution, we fit them with a linear combination of the Gaussian distribution and the exponential decaying function. They are fitted very well with this function. Then, the areas of the Gaussian part and the exponential decaying part are plotted in Fig.~\ref{fig:fig3}-(c),(d). This shows which contribution is dominant to the distribution.
At shorter polymer length ($N_p = 75, 200$), the exponential decaying part is dominant regardless of $k_\theta$. This is because of the artificial retracting force from the quenched end points. The artificial effect from end points gets weaker at larger chain length.
In polymer lengths of $N_p = 300\sim 1000$, the trends are clearly splitted as $k_\theta$. At $k_\theta = 0$ (flexible polymer with larger $N_e$), the Gaussian part is dominant. The trend is a bit dependent on the chain length, and the Gaussian part is slightly more important for longer chains. This can be understood from the weaker contribution from the fixed ends for longer chains.
In contrast, the exponential decaying part is more important for $k_\theta = 2$. The contribution is larger for longer chains.
This is because longer chains can form tighter networks, and thus the interaction energy at intersections is stronger. Indeed as we expected, the interaction energy at the intersections and the linking number increase the mass term.


\underline{{Towards the second ``quantization"}:}
Our phenomenological Maxwell-London equation based on the topological BF theory seems to be not inconsistent with the successful tube model. However, there still exist important unanswered questions, which may be difficult to be solved within the present formulation. For example, it is not clear how to express the mass parameter as a function of measurable physical quantities. We point out that this fundamental difficulty of our formulation results from the fact that the theory is written in the language of the first ``quantization". We need to represent the BF theory in the second-quantization formulation, where not only gauge dynamics but also polymer dynamics is expressed in terms of field variables instead of the position variable. This work has been performed before~\cite{matsen2006self}, where the generating function of
\begin{widetext}
\bqa && W = \int D q(\mathbf{r},s) \exp\Bigl\{- \int_{0}^{N} d s \int d^{3} \mathbf{r} \frac{1}{2} q(\mathbf{r},s) ( \partial_{s} - D \mathbf{\nabla}_{\mathbf{r}}^{2} ) q(\mathbf{r},s) \Bigr\} \eqa
\end{widetext}
leads to the Edwards-Anderson equation \bqa && \partial_{s} q(\mathbf{r},s) = D \mathbf{\nabla}_{\mathbf{r}}^{2} q(\mathbf{r},s) \eqa in the saddle-point analysis and an internal energy \bqa && E = - \lim_{N \rightarrow \infty} \frac{1}{N} \ln W . \eqa The density and current of polymers are given by \bqa && \rho(\mathbf{r},s) = q^{\dagger}(\mathbf{r},s) q(\mathbf{r},s) \eqa and \bqa \mathbf{j}(\mathbf{r},s) = - \frac{1}{2D} \Bigl( q^{\dagger}(\mathbf{r},s) [\mathbf{\nabla}_{\mathbf{r}} q(\mathbf{r},s)] - [\mathbf{\nabla}_{\mathbf{r}} q^{\dagger}(\mathbf{r},s)] q(\mathbf{r},s) \Bigr) , \nn \eqa respectively, which satisfy the conservation law \bqa && \partial_{s} \rho(\mathbf{r},s) + \mathbf{\nabla}_{\mathbf{r}} \cdot \mathbf{j}(\mathbf{r},s) = 0 . \eqa Here, the $q^{\dagger}(\mathbf{r},s)$ field satisfies \bqa && \partial_{s} q^{\dagger}(\mathbf{r},s) = - D \mathbf{\nabla}_{\mathbf{r}}^{2} q^{\dagger}(\mathbf{r},s) . \eqa

Since the equation of motion for the $q(\mathbf{r},s)$ field differs from that of the $q^{\dagger}(\mathbf{r},s)$ field, it is convenient to introduce a Nambu-spinor~\cite{bardeen1957theory} \bqa && \psi(\mathbf{r},s) = \left(\begin{array}{cc} q(\mathbf{r},s) \\ q^{\dagger}(\mathbf{r},s) \end{array} \right) . \eqa Then, the generating function can be reformulated as follows
\begin{widetext}
\bqa && W = \int D \psi^{\dagger}(\mathbf{r},s) D \psi(\mathbf{r},s) \exp\Bigl\{- \int_{0}^{N} d s \int d^{3} \mathbf{r} \frac{1}{4} \psi^{\dagger}(\mathbf{r},s) ( \partial_{s} \mathbf{\tau}_{3} - D \mathbf{\nabla}_{\mathbf{r}}^{2} ) \psi(\mathbf{r},s) \Bigr\} , \eqa
\end{widetext}
where the density and current of polymers are expressed by \bqa && \rho(\mathbf{r},s) = \frac{1}{2} \psi^{\dagger}(\mathbf{r},s) \psi(\mathbf{r},s) \eqa
and \bqa \mathbf{j}(\mathbf{r},s) = - \frac{1}{4D} \Bigl( \psi^{\dagger}(\mathbf{r},s) \mathbf{\tau}_{3} [\mathbf{\nabla}_{\mathbf{r}} \psi(\mathbf{r},s)] - [\mathbf{\nabla}_{\mathbf{r}} \psi^{\dagger}(\mathbf{r},s)] \mathbf{\tau}_{3} \psi(\mathbf{r},s) \Bigr) \nn && \eqa respectively.

Introducing the chemical potential ($\mu$) and effective interactions ($V$) of polymers, we can describe interacting polymers at finite density, given by
\begin{widetext}
\bqa && W = \int D \psi^{\dagger}(\mathbf{r},s) D \psi(\mathbf{r},s) \exp\Bigl[- \int_{0}^{N} d s \int d^{3} \mathbf{r} \Bigl\{ \frac{1}{4} \psi^{\dagger}(\mathbf{r},s) ( \partial_{s} \mathbf{\tau}_{3} - \mu - D \mathbf{\nabla}_{\mathbf{r}}^{2} ) \psi(\mathbf{r},s) + \frac{V}{4} [\psi^{\dagger}(\mathbf{r},s) \psi(\mathbf{r},s)]^{2} \Bigr\} \Bigr] . \eqa
\end{widetext}
Entanglement of polymers can be formulated as before, introducing gauge fields as follows
\begin{widetext}
\bqa && W = \sum_{N_{G} = 0 }^{\infty} \frac{1}{N_{G}!} \int d \kappa e^{i \kappa N_{G}} \int D A_{\mu}(\mathbf{r},s) \int D \psi^{\dagger}(\mathbf{r},s) D \psi(\mathbf{r},s) \nn && \exp\Bigl[- \int_{0}^{N} d s \int d^{3} \mathbf{r} \Bigl\{ \frac{1}{4} \psi^{\dagger}(\mathbf{r},s) \Bigl( [\partial_{s} - i \kappa A_{s}(\mathbf{r},s) \mathbf{\tau}_{3}] \mathbf{\tau}_{3} - \mu - D [\mathbf{\nabla}_{\mathbf{r}} - i \kappa \mathbf{A}(\mathbf{r},s) \mathbf{\tau}_{3}]^{2} \Bigr) \psi(\mathbf{r},s) \nn && - \kappa^{2} \frac{D}{4} \psi^{\dagger}(\mathbf{r},s) \psi(\mathbf{r},s) [\mathbf{A}_{\mathbf{r}}(\mathbf{r},s)]^{2} + \frac{V}{4} [\psi^{\dagger}(\mathbf{r},s) \psi(\mathbf{r},s)]^{2} + i \frac{\theta}{2\pi} \epsilon_{\mu\nu\lambda} A_{\mu}(\mathbf{r},s) \partial_{\nu} A_{\lambda}(\mathbf{r},s) \Bigr\} \Bigr] . \eqa
\end{widetext}
Here, $\kappa$ is the Lagrange multiplier field to impose the Gaussian linking number and $N_{G}$ is the total number of the Gaussian linking number. In this formulation we allow self-linking of a polymer in addition to mutual entanglement, described by the Chern-Simons term \bqa && \mathcal{S}_{CS} = \int_{0}^{N} d s \int d^{3} \mathbf{r} \Bigl( i \frac{\theta}{2\pi} \epsilon_{\mu\nu\lambda} A_{\mu}(\mathbf{r},s) \partial_{\nu} A_{\lambda}(\mathbf{r},s) \Bigr) , \nn \eqa where $\theta$ is a statistical angle, which can be used as a phenomenological parameter. By substituting the time derivative $\partial_{t}$ into the segment derivative  $\partial_{s}$, we can revisit the previous effective Maxwell-London equation from the effective Chern-Simons field theory.

This field-theoretic formulation allows us to introduce renormalization effects of both dynamics of gauge and polymer fields self-consistently at least in the level of random phase approximation (RPA)~\cite{anderson1958random}. Although this Chern-Simons field theory has a similar structure with that of fractional quantum Hall effect, we can't apply its solution directly. There exists a substantial difference between two Chern-Simons field theories: The time derivative does not have an $i$ factor, where $i$ is a complex number with $i^{2} = - 1$~\cite{note1}. Self-consistent renormalizations within the RPA level are required near future.

\underline{{Discussion}:}
Starting from the topological BF theory in the first-quantization representation, we derive the effective Maxwell-London equation. We connect the solution of this phenomenological equation with the physics of the tube model. The essence of the tube model is explained by the Meissner effect of our phenomenological equation. We revisit this formula with an alternative approach based on second quantization. These results seem to be consistent with our numerical analysis.

We would like to emphasize that the role of the topological constraint (the preserved Gaussian linking number) differs from that of effective interactions between segments of polymers. Effective interactions between segments of polymers can be translated into dynamics of longitudinal gauge fluctuations, \textit{i.e.}, Coulomb interactions, which would be screened by polymer fluctuations, nothing but the Debye screening in metals~\cite{altland2010condensed} and thus, allowed to be neglected at low energies. This massive dynamics of longitudinal gauge fluctuations should be distinguished from Higgs phenomena, where transverse gauge fluctuations become massive, realized in the SC media. The phenomenological Maxwell-London equation implies that the tube model should be understood within the presence of the topological constraint beyond effective interactions between polymer segments. 

The Chern-Simons field theory allows us to pursue an analogy between superconductivity and entangled polymer complex in depth. The fundamental concept of U(1) symmetry breaking in superconductivity implies the gapless Goldstone mode and gapped Higgs mode, where the gapless sound mode is pushed up to the plasmon mode when there exist transverse gauge fluctuations~\cite{altland2010condensed}. In order to search such plasmon and Higgs modes in the entangled polymer complex, we need to investigate the dynamics of entangled polymers, integrating over gauge fluctuations. This deep connection can be realized in the second-quantization representation. Then, the dynamics of corresponding the plasmon and Higgs modes would reveal characteristic responses of the entangled polymer complex such as compressibility, viscosity, and etc. These will be our future direction.

\acknowledgements
KS was supported by the Ministry of Education, Science, and Technology
(No. 2012R1A1B3000550 and No. 2011-0030785) of the National Research Foundation
of Korea (NRF) and by TJ Park Science Fellowship of the POSCO TJ Park
Foundation.YSJ was supported by the Ministry of Education, Science, and Technology (NRF-2012R1A1A2009275, NRF-C1ABA001-2011-0029960) of the National Research Foundation
of Korea (NRF).
\bibliography{refs}

\begin{thebibliography}{33}
\bibitem[{note1()}]{note1}
note1, \bibinfo{note}{recall that the same factor of $i$ difference in the
  time-derivative term appears in analogy between the Gaussian chain under the
  external field and the Schrodinger equation.}
\end{thebibliography}

\end{document}